\documentclass[12pt]{iopart}

\usepackage{graphicx,amssymb}

\begin{document}

\title{Constraints on the generalized natural inflation after Planck 2018}

\author{Nan Zhang$^1$, Ya-Bo Wu$^1$, Jun-Wang Lu$^2$, Chu-Wen Sun$^1$, Li-Jie Shou$^1$, Hai-Zhou Xu$^1$ }

\address{$^1$ Department of Physics, Liaoning Normal University, Dalian 116029, P.R.China}
\address{$^2$ School of Physics and Electronics, Qiannan Normal University for Nationalities, Duyun 558000, P.R.China}
\ead{ybwu61@163.com}
\vspace{10pt}


\begin{abstract}
Based on the dynamics of single scalar field slow-roll inflation and the theory of reheating, we investigate the generalized natural inflationary (GNI) model. Concretely, we give constraints on the scalar spectral index $n_{s}$ and tensor-to scalar ratio $r$ for $\Lambda$CDM $+r$ model according to the latest data from Plack 2018 TT,TE,EE+lowE+lensing (P18) and BICEP2/Keck 2015 season (BK15), i.e., $n_{s}=0.9659\pm0.0044$ at $68\%$ confidence level (CL) and $r<0.0623$ at $95\%$CL. We find that the GNI model is favored by P18 plus BK15 in the ranges of $\log_{10}(f/M_{p})=0.62^{+0.17}_{-0.18}$ and $m=0.35^{+0.13}_{-0.23}$ at $68\%$CL.
In addition, the corresponding predictions of the general and two-phase reheating are respectively discussed. It follows that the parameter $m$ has the significant effect on the model behaviors.
\end{abstract}

\section{\label{sec:level11}Introduction}

Inflation is a widely accepted supplement to the successful standard big bang theory.
The existence of inflation phase can lead to a rapid accelerated expansion period of the universe in order to solve the problems such as the flatness, the horizon, etc \cite{Guth1981,Linde1982,AandS1982,Hawking1982B110,Linde1983}. It can also give a superior interpretation of the origin of structure and cosmic microwave background (CMB) \cite{MandC1981,Hawking1982B115,GandP1982}.
Then the cold and empty universe during inflation is heated through the reheating phase, the radiation particles corresponding to the standard model are also generated in the reheating epoch.
In order to investigate the properties of inflation period, many kinds of models have been proposed,
such as the $R^{2}$ model \cite{Starobinsky1980}, hilltop model \cite{BandL2005}, natural inflation model \cite{Fet1990,Aet1993}, $\alpha$ attractors \cite{KandL2013,Ket2013} and so on.
Most models are slow-roll ones taking a scalar field as the inflaton and making it slowly roll toward its true ground state \cite{LandL1992,Let1994,SandT1984}.

As we know, the natural inflationary (NI) model is a kind of single field slow-roll inflationary models.
It was proposed in Refs.~{\cite{Fet1990}} with the potential form $V(\phi)=\Lambda^{4}[1\pm cos(N\phi/f)]$, in which the choice of sign has no effect on the results, and usually, taking $N=1$.
The model parameter $f$ is called the decay constant \cite{MandK2015} and $f\gtrsim0.3M_{p}$ \cite{Aet1993}.
The NI model is widely studied in many literatures because of its simple potential form and theoretically well motivation \cite{ni1,ni2,ni3}.
However, from the aspect of recent observation, the NI model is disfavored by the data from Planck 2018 \cite{Pgroup2014A16,Pgroup2014A22,Pgroup2016A13,Pgroup2016A20,Pgroup2018P,Pgroup2018X} and BICEP2/Keck with inclusion of 95 GHz band (BK14) \cite{PandBgroup2016}.

The ``generalized'' version of the NI model was proposed in Ref.~{\cite{MandK2015}}, which adds the model parameter $m$ on the basis of the NI model, i.e., $V(\phi) = 2^{1-m}\Lambda^{4}[1+cos(\frac{\phi}{f})]^{m}$ \cite{MandK2015}.
Here, we call it the generalized natural inflationary (GNI) model.
Evidently, the NI model is a special case of the GNI model corresponding to $m=1$.
Moreover, for the investigation of the GNI model in Ref.~{\cite{MandK2015}}, the authors discussed only the proper ranges of values of $N_{*}$ by using the tight constraint of $0\lesssim w_{re}\lesssim0.25$ \cite{Pet2006} and taking $m=1$ in the most cases.
Whereas the corresponding results to the cases of $m\neq1$ have been rarely discussed.
Thus, it is an interesting issue that how the parameter $m$ influences the behaviors of the GNI model.
In addition, we are inspired to wonder if the GNI model, which is the broad class of the NI model, could be favored by the recent data from Plack 2018 TT,TE,EE+lowE+lensing (P18) and BICEP2/Keck 2015 season (BK15) \cite{BK15} due to the existence of parameter $m$.
These are just our motivations and purposes of investigating the GNI model in this paper.

Based on the above, this paper will focus on the key parameters for the inflationary models, i.e., scalar spectral index $n_{s}$, tensor-to-scalar ratio $r$, the e-folding number $N_{*}$, the reheating e-folding number $N_{re}$, the reheating temperature $T_{re}$ and the effective average equation of state (EoS) $w_{re}$ \cite{Aet1982,Ket1997,Aet2010}, etc.
Specifically, we will investigate the constraints on $n_{s}$ and $r$ by means of the public codes Cosmomc \cite{cosmomc} according to the data from P18 plus BK15, as well as the allowable parameter space of the GNI model, in which we will calculate the running spectral index $\alpha_{s}$.
In addition, we will study two different mechanisms of the reheating phase, one is the general reheating phase, and the other is a two-phase reheating process \cite{MD2016,UandY2016}.
Our research results indicate that the GNI model is favored by P18 plus BK15 in the ranges of $\log_{10}(f/M_{p})=0.62^{+0.17}_{-0.18}$ and $m=0.35^{+0.13}_{-0.23}$ at $68\%$CL.
Moreover, the parameter $m$ has the significant effect on the model behaviors.
The evolutions of the reheating parameters of the GNI model are also discussed in detail, including $N_{re}$, $T_{re}$ in the general reheating phase and the oscillation e-folding number $N_{sc}$, temperature $T_{re}e^{N_{th}}$, coupling constant $g$ in the two-phase reheating.

This paper is organized as follows.
In Sec.~\ref{sec:level1}, we briefly review the depiction of single scalar field slow-roll inflation, the constraints on $n_{s}$ and $r$ from the data of P18 plus BK15 are obtained.
In Sec.~\ref{sec:level2}, we investigate the validity of the GNI model, i.e., the constraints on the model parameters according to the observational data.
Then in Sec.~\ref{sec:level3}, we discuss the general reheating phase and the two-phase reheating of the GNI model.
Sec. \ref{sec:level14} presents the conclusions.

\section{\label{sec:level1}The single field slow-roll inflation}

We will give a brief review of a single field inflationary model \cite{ni1,ni2,Pgroup2014A22,Pgroup2016A20,Pgroup2018X,UandY2016,Cheng:2014bta,Huang:2015gca} and we start with the equations of motion induced by a scalar field $\phi$ in the frame of spatially flat FRW background universe,
\begin{equation}\label{eq:F}
H^{2}=\frac{1}{3M_{p}^{2}}(\frac{1}{2}\dot{\phi}+V(\phi))\mbox{,}
\end{equation}
\begin{equation}\label{eq:KG}
\ddot{\phi}+3H\dot{\phi}=-V'(\phi)\mbox{,}
\end{equation}
where $H\equiv\dot{a}/a$ is the Hubble parameter, $V(\phi)$ is the potential of field $\phi$, $M_{p}\equiv\frac{1}{\sqrt{8\pi G}}\simeq2.435\times10^{18}$ GeV is the reduced Planck mass, the dot denotes differentiation with respect to cosmic time $t$ and the prime denotes differentiation with respect to $\phi$.

In the case of slow-roll inflation, the potential term dominates the total energy density and the scalar field changes slowly with time, Eqs. (\ref{eq:F}) and (\ref{eq:KG}) can be written as follows:
\begin{equation}\label{eq:Fsr}
H^{2}\simeq\frac{V(\phi)}{3M_{p}^{2}}\mbox{,}
\end{equation}
\begin{equation}\label{eq:KGsr}
3H\dot{\phi}\simeq -V'(\phi)\mbox{.}
\end{equation}

Thus, the parameter $N_{*}$, which represents the e-folding number between the pivot scale $k_{*}$ exiting from the Hubble radius and the end of inflation, can be expressed in terms of the potential $V(\phi)$ under the slow-roll approximation
\begin{equation}\label{eq:Nk}
N_{*}\equiv\ln\frac{a_{end}}{a_{*}}=\int_{^{t_{*}}}^{t_{end}}Hdt
\simeq-\frac{1}{M_{p}^{2}}\int_{^{\phi_{*}}}^{\phi_{end}}\frac{V(\phi)}{V'(\phi)}d\phi\mbox{,}
\end{equation}
where the subscripts ``$*$'' and ``$end$'' correspond to crossing the horizon and the end of inflation, respectively.

Next, introducing the slow-roll parameters:
\begin{equation}\label{eq:ev1}
\epsilon_{v}=\frac{M_{p}^{2}}{2}\frac{V'(\phi)^{2}}{V(\phi)^{2}}\mbox{,}
\end{equation}
\begin{equation}\label{eq:ev2}
\eta_{v}=M_{p}^{2}\frac{V''(\phi)}{V(\phi)}\mbox{,}
\end{equation}
The power spectra of curvature and tensor perturbations $\mathcal{P}_{\mathcal{R}}$, $\mathcal{P}_{t}$ can be well approximated in the case of the usual single field slow-roll inflationary models, thus the scalar spectral index $n_{s}$ and the tensor spectral index $n_{t}$ can be expressed as:
\begin{equation}\label{eq:ns}
n_{s}\simeq 1-6\epsilon_{v}+2\eta_{v}\mbox{,}
\end{equation}
\begin{equation}\label{eq:nt}
n_{t}\simeq -2\epsilon_{v}\mbox{.}
\end{equation}
Then making use of the scalar power spectra amplitude $A_{s}\simeq\frac{V}{24\pi^{2}M_{p}^{4}\epsilon_{v}}$ and the tensor amplitude $A_{t}\simeq\frac{2V}{3\pi^{2}M_{p}^{4}}$, the tensor-to-scalar ratio $r$ can be obtained as
\begin{equation}\label{eq:r}
r=\frac{A_{t}}{A_{s}}\simeq 16\epsilon_{v}\mbox{,}
\end{equation}
which means $n_{t}$ is not a free parameter due to $r\simeq-8n_{t}$.

\begin{figure}[htbp]
\centering
\includegraphics[width=8cm]{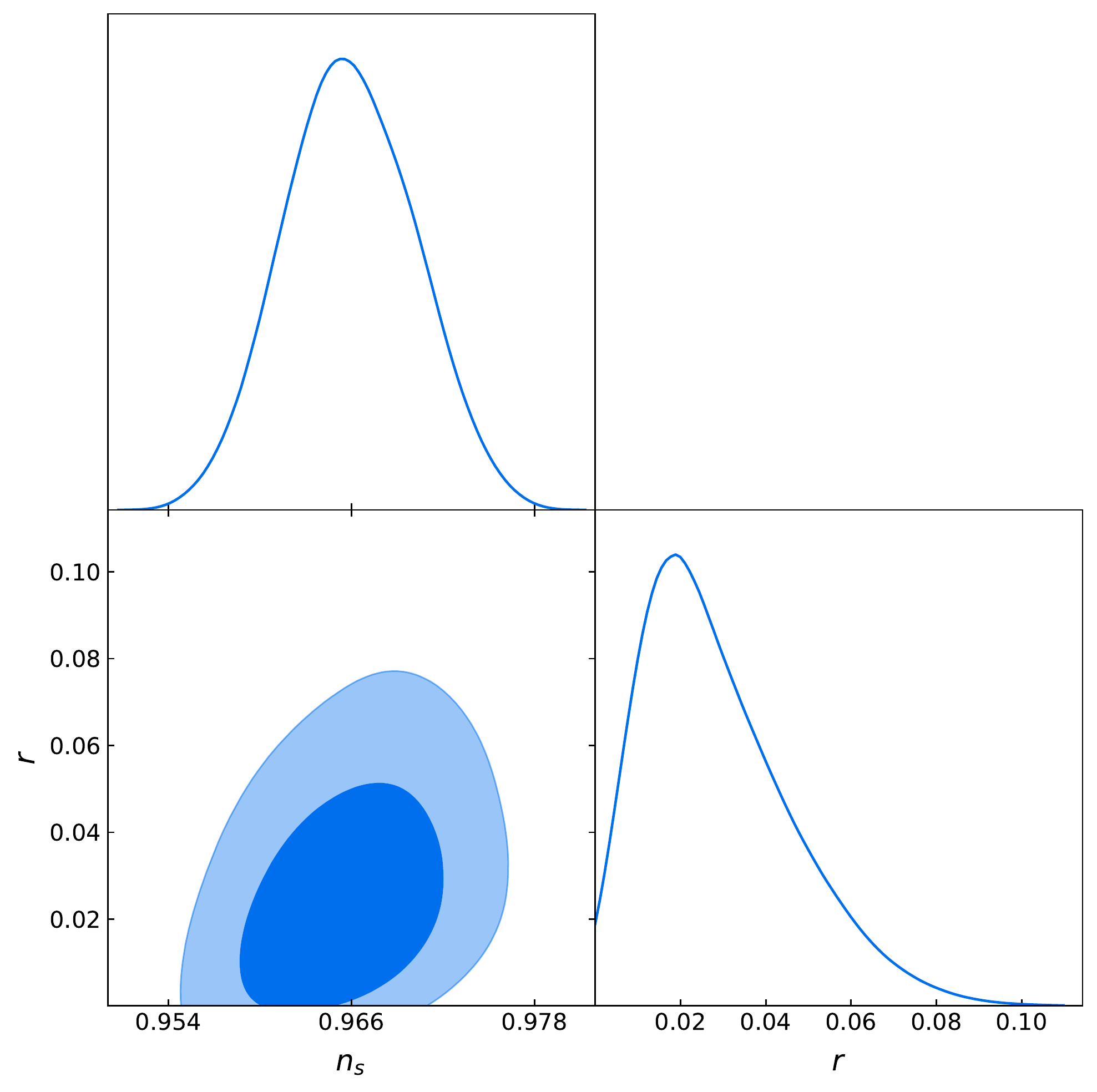}
\caption{\label{fig:data}The marginalized contour plots and likelihood distributions for $n_{s}$ and $r$ at 68\%CL and 95\%CL from the data of P18+BK15, respectively.}
\end{figure}

In this paper, we adopt the data from P18 \cite{Pgroup2018X} plus BK15 \cite{BK15} to obtain the constraints on $n_{s}$ and $r$ for $\Lambda$CDM $+r$ model, they are given as follows by means of the available Cosmomc code \cite{cosmomc}:
\begin{equation}\label{eq:datans}
n_{s}=0.9659\pm0.0044~~(68\%CL)\mbox{,}
\end{equation}
\begin{equation}\label{eq:datar}
r<0.0623~~(95\%CL)\mbox{.}
\end{equation}
The contour plots for $n_{s}$ and $r$ are shown in Figure~\ref{fig:data}. It indicates that the power spectrum of curvature perturbation deviates from the exact scale-invariant power spectrum at more than $7\sigma$ CL.

\section{\label{sec:level2}The constraints on the model parameters of the GNI model}

\begin{figure}[htbp]
\centering
\includegraphics[width=6cm]{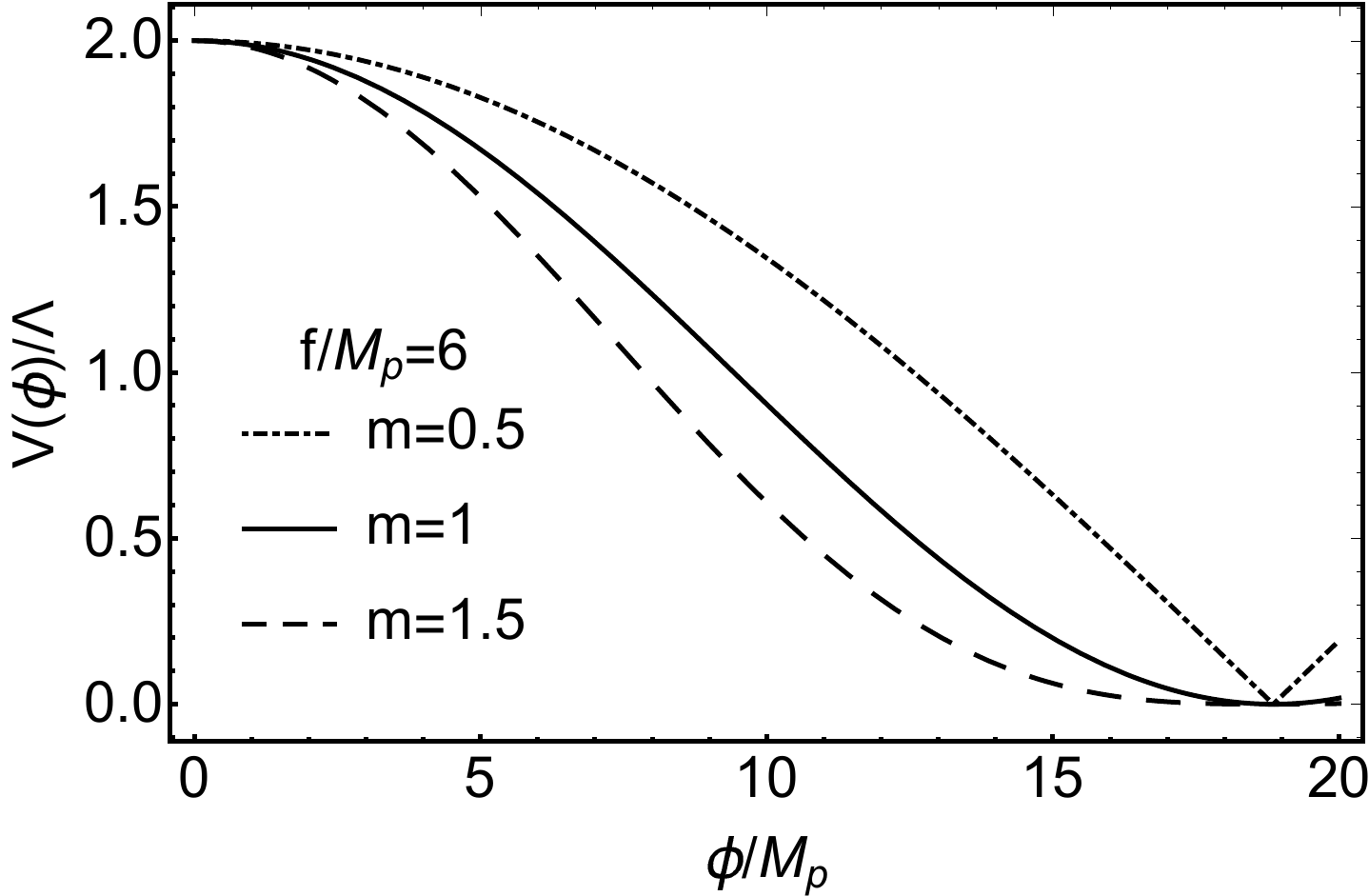}
\caption{\label{fig:V} The evolving trajectories of $V(\phi)$ in the GNI model for $m=0.5, 1$ and $1.5$, respectively. Here, $f/M_{p}=6$ is taken.}
\end{figure}

As we know, the ordinary NI potential was simply generalized by adding one parameter $m$ in Ref. {\cite{MandK2015}},
here we call this model as the generalized natural inflationary (GNI) model.
The form of potential for the GNI model is expressed as follows:
\begin{equation}\label{eq:GNIV}
V(\phi) = 2^{1-m}\Lambda^{4}[1+cos(\frac{\phi}{f})]^{m}\mbox{,}
\end{equation}
where the energy density $\Lambda^{4}$, decay constant $f$ and the constant $m$ are the parameters of the model.
It follows that when $m=1$, it can reduce to the so-called NI model.
If $f\rightarrow\infty$, the NI model seems to behave like the chaotic inflationary model.
Similarly, the GNI model behaves as a pure power law model when $f\rightarrow\infty$ \cite{MandK2015}.
Figure~\ref{fig:V}
shows the evolving tendency of the potential roughly. The horizontal axis is $\phi/M_{p}$, the vertical axis is $V(\phi)/\Lambda^{4}$.
From the trajectories, we can find that if $f$ is a fixed value, such as $f/M_{p}=6$, $V(\phi)$ changes more slowly for smaller $m$ in the beginning and becomes steep in the end, the trajectories reach zero at the same value of $\phi$.

According to Eqs. (\ref{eq:ev1}) and (\ref{eq:ev2}), the slow-roll parameters $\epsilon_{v}$ and $\eta_{v}$ of the GNI model can be given as follows:
\begin{equation}\label{eq:GNIev1}
\epsilon_{v} =\frac{M_{p}^{2}m^{2}}{2f^{2}}[\frac{1-cos(\phi/f)}{1+cos(\phi/f)}]\mbox{,}
\end{equation}
\begin{equation}\label{eq:GNIev2}
\eta_{v} =-\frac{M_{p}^{2}}{f^{2}}[\frac{m-m^{2}(1-cos(\phi/f))}{1+cos(\phi/f)}]\mbox{.}
\end{equation}

\begin{figure}[htbp]
\centering
\includegraphics[width=7cm]{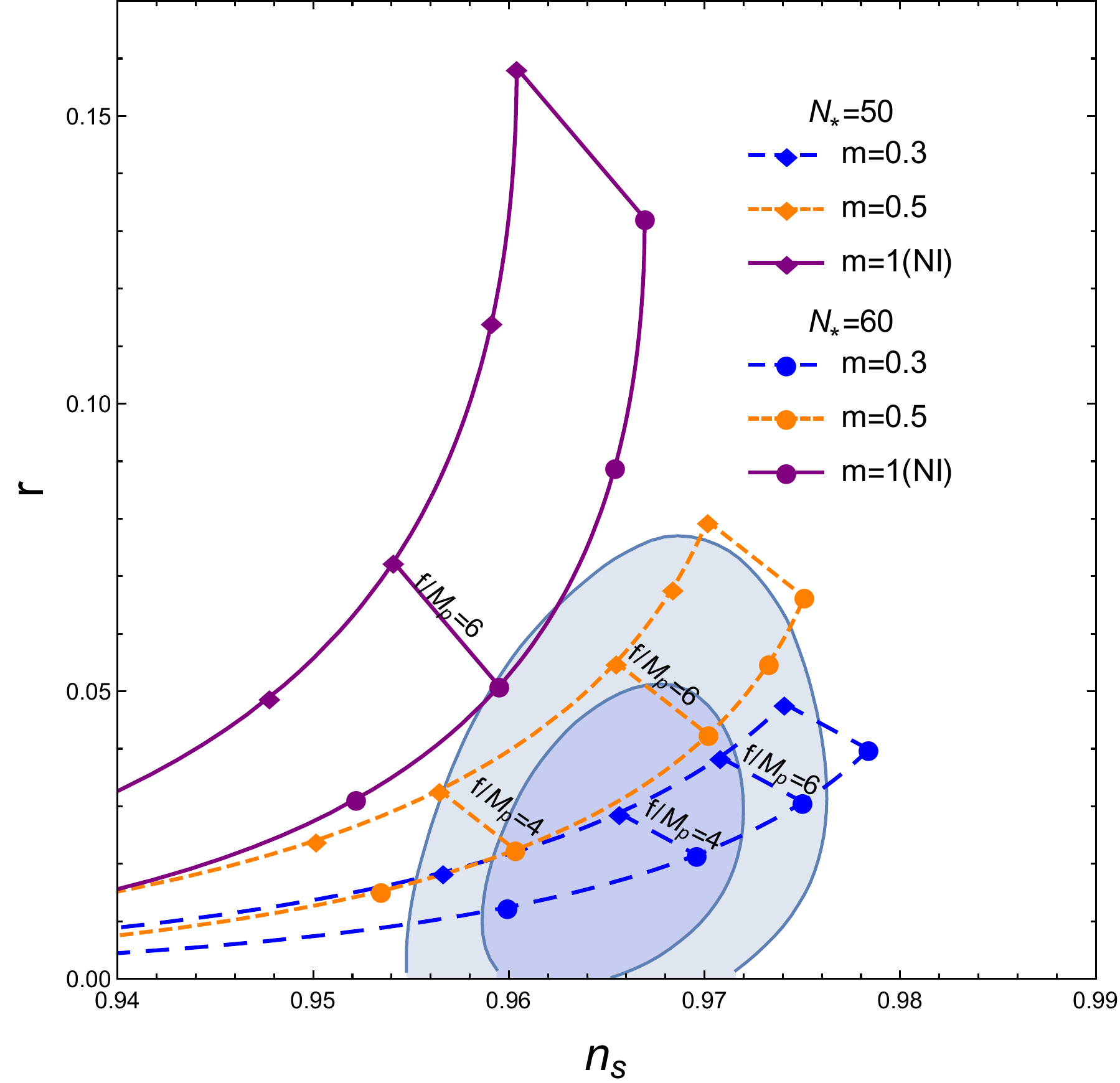}
\caption{\label{fig:nsr} The plots of $r$ and $n_{s}$ in the GNI model for $m=0.3, 0.5$ and $1$, respectively. $N_{*}\in[50, 60]$ is taken and the shaded regions represent the constraints given by P18+BK15 at $68\%$ and $95\%$ CL, respectively.}
\end{figure}

Thus, $n_{s}$ and $r$ can be expressed as:
\begin{equation}\label{eq:GNIns}
n_{s} =1-\frac{M_{p}^{2}}{f^{2}}[\frac{m^{2}(1-cos(\phi/f)+2m)}{1+cos(\phi/f)}]\mbox{,}
\end{equation}
\begin{equation}\label{eq:GNIr}
r =\frac{8M_{p}^{2}m^{2}}{f^{2}}[\frac{1-cos(\phi/f)}{1+cos(\phi/f)}]\mbox{.}
\end{equation}
And the e-folding number $N_{*}$ is derived as:
\begin{equation}\label{eq:GNINk}
N_{*}=\frac{f^{2}}{mM_{p}^{2}}\ln\frac{1-cos(\phi_{end}/f)}{1-cos(\phi_{*}/f)}\mbox{.}
\end{equation}
When $\epsilon_{v}=\epsilon_{end}=1$, $\phi_{end}$ can be obtained from Eq. (\ref{eq:GNIev1}) as follows
\begin{equation}\label{eq:phiEnd}
\phi_{end}=f\arccos\frac{m^{2}M_{p}^{2}-2f^{2}}{m^{2}M_{p}^{2}+2f^{2}}\mbox{.}
\end{equation}
Substituting Eq. (\ref{eq:phiEnd}) into Eq. (\ref{eq:GNINk}), $\phi_{*}$ can be derived as
\begin{equation}\label{eq:phik}
\phi_{*}=f\arccos[1-\frac{4f^{2}}{m^{2}M_{p}^{2}+2f^{2}}\exp(-\frac{mM_{p}^{2}}{f^{2}}N_{*})]\mbox{.}
\end{equation}
It follows that $\phi_{*}$ is the function of $(N_{*}, f, m)$. In this case, Eqs. (\ref{eq:GNIns}) and (\ref{eq:GNIr}) become into
\begin{equation}\label{eq:GNInsN}
n_{s} =1-\frac{mM_{p}^{2}}{f^{2}}[1+\frac{2f^{2}(m+1)\exp(-\frac{mM_{p}^{2}}{f^{2}}N_{*})}{m^{2}M_{p}^{2}+2f^{2}(1-\exp(-\frac{mM_{p}^{2}}{f^{2}}N_{*}))}]\mbox{,}
\end{equation}
\begin{equation}\label{eq:GNIrN}
r =\frac{16m^{2}M_{p}^{2}\exp(-\frac{mM_{p}^{2}}{f^{2}}N_{*})}{m^{2}M_{p}^{2}+2f^{2}(1-\exp(-\frac{mM_{p}^{2}}{f^{2}}N_{*}))}\mbox{.}
\end{equation}
It is easy to see that when taking $m=1$, Eqs. (\ref{eq:GNInsN}) and (\ref{eq:GNIrN}) can reduce to the results in the NI model.
Hence, the predictions of $n_{s}$ and $r$ in the GNI model are given in Figure~\ref{fig:nsr}, $N_{*}$ is taken in the usual range of $[50, 60]$, and the shaded regions represent the constraints given by P18+BK15 at $68\%$ and $95\%$ CL, respectively.
It can be found that the case of $m=1$ (NI model) is disfavored by the data of P18 plus BK15, however, the case of $m<1$ can provide the small values of $r$ in the proper ranges of values of $n_{s}$.
Thus, the GNI model is worth investigating in the case of $m<1$.

\begin{figure}[htbp]
\centering
\includegraphics[width=4.3cm]{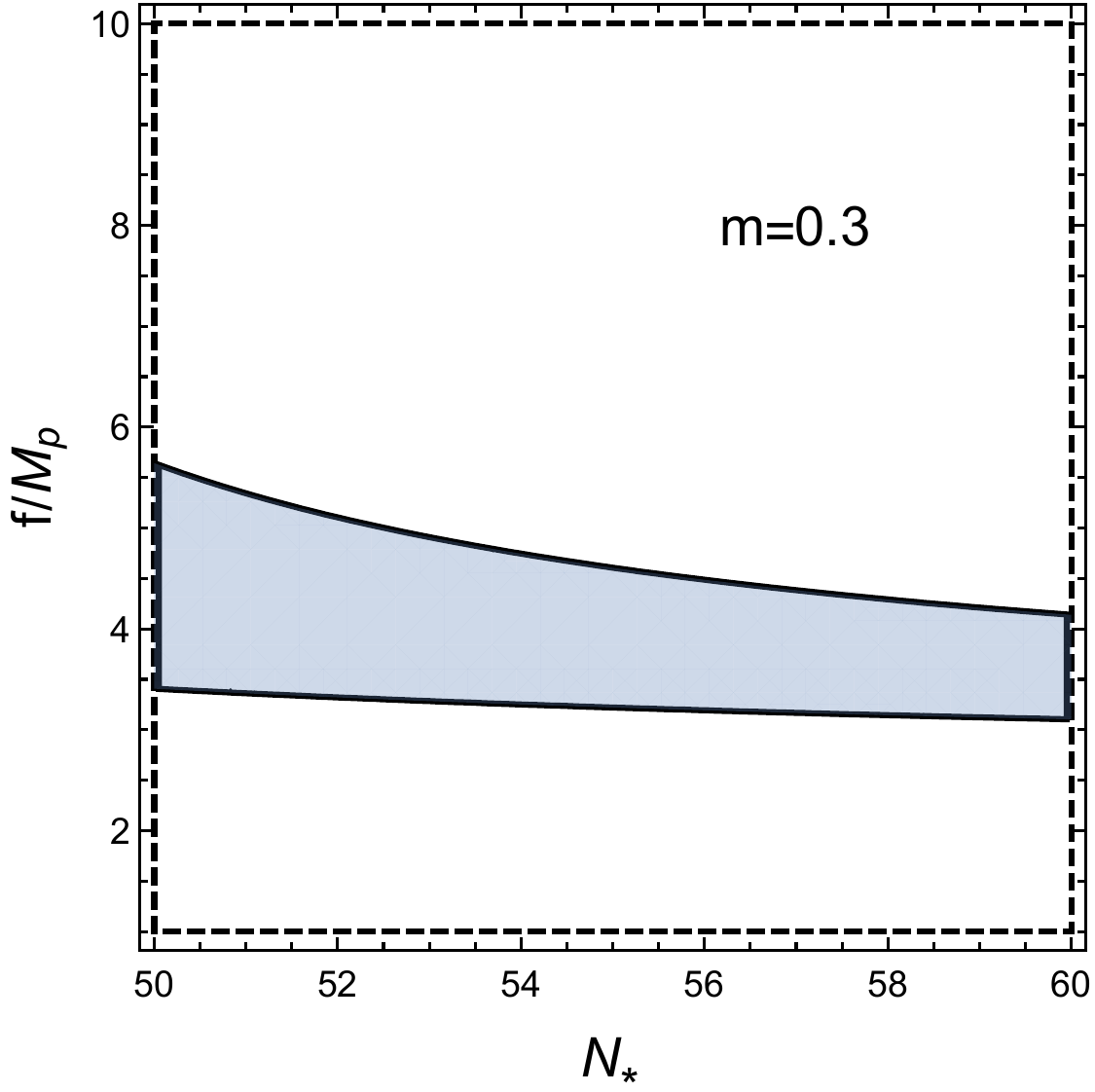}
\includegraphics[width=4.3cm]{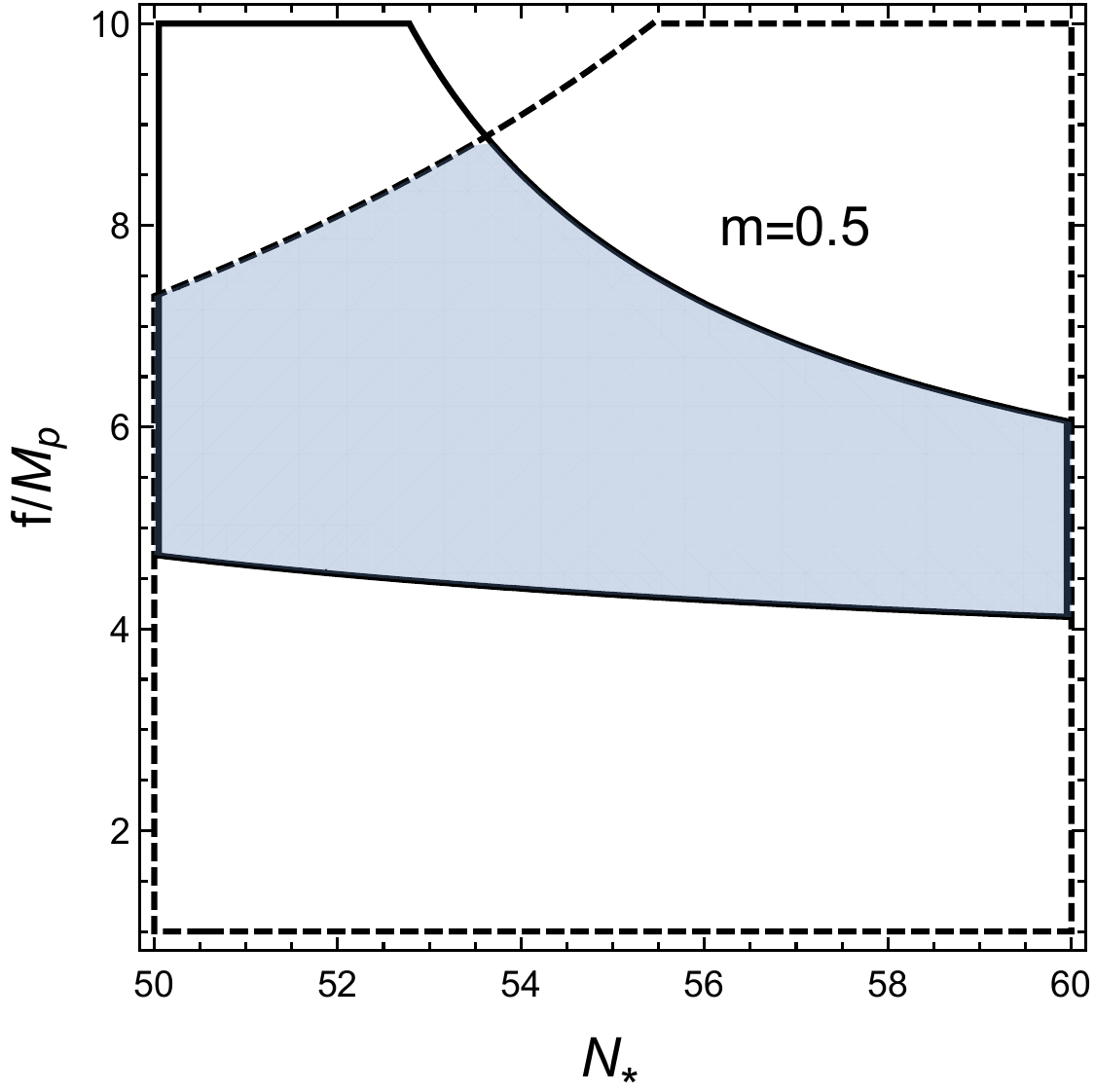}
\includegraphics[width=4.3cm]{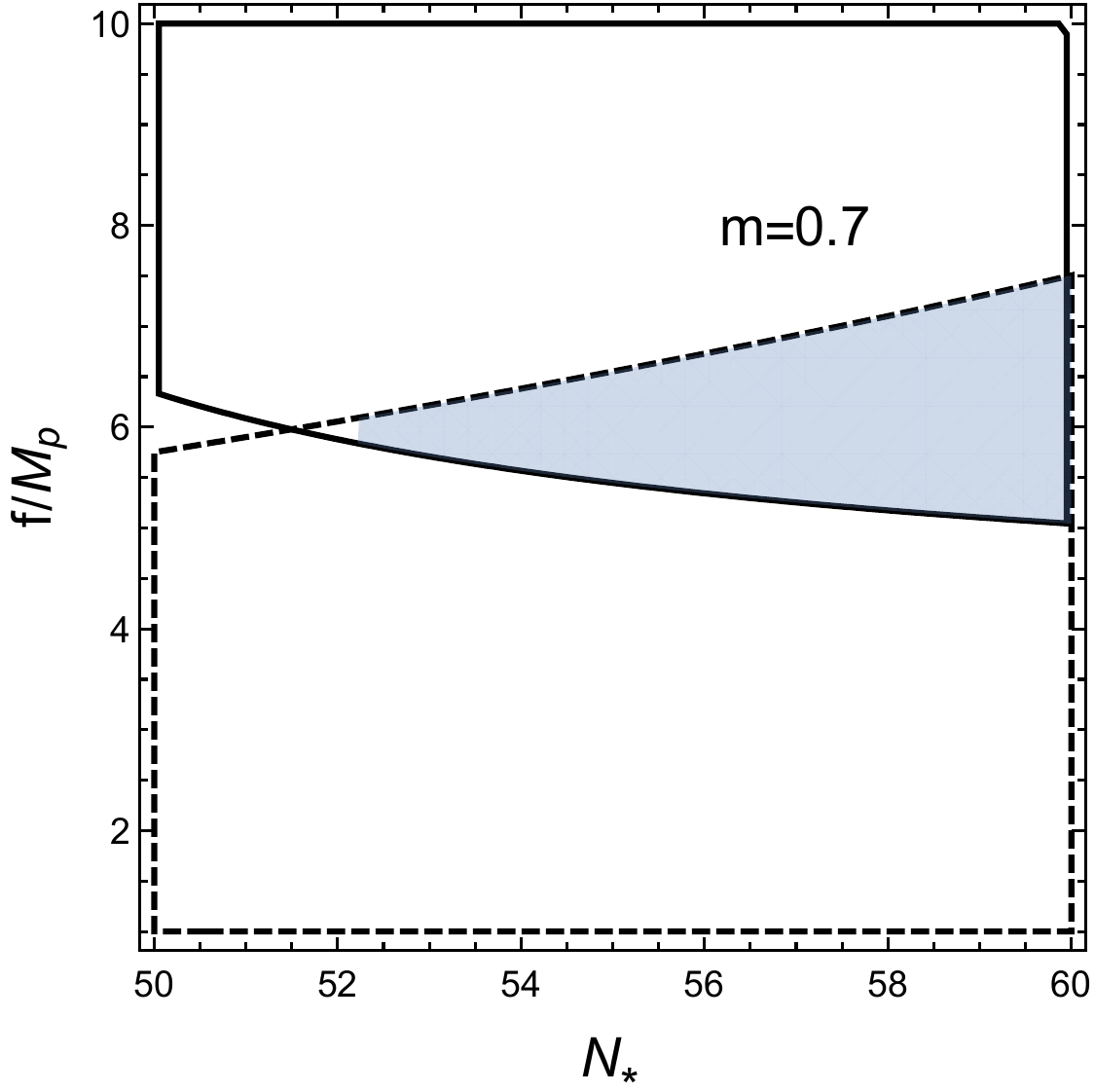}
\caption{Allowed ranges of values of $N_{*}$ and $f$ for $m=0.3$, $0.5$ and $0.7$, respectively. In each panel, the solid curves bracket the range of values of $n_{s}=0.9659\pm0.0044$, the dotted curves bracket the range of values of $r<0.0623$.}
\label{fig:nsrf} 
\end{figure}
Figure~\ref{fig:nsrf} shows the allowed ranges of values of $N_{*}$ and $f$ when taking $m=0.3$, $0.5$ and $0.7$, respectively.
It can be seen that the allowable parameter space firstly becomes large and then becomes small with increasing $m$ according to the areas of the shaded regions in Figure~\ref{fig:nsrf}.
Moreover, when $m=0.7$, $N_{*}^{min}=51.1$ can be obtained, which suggests that the minimum value of $N_{*}$ could be larger than $50$ if the value of $m$ is relatively large enough.
Besides, the value of $f/M_{p}$ is smaller than $10$ from Figure~\ref{fig:nsrf}.

\begin{figure}[htbp]
\centering
\includegraphics[width=8cm]{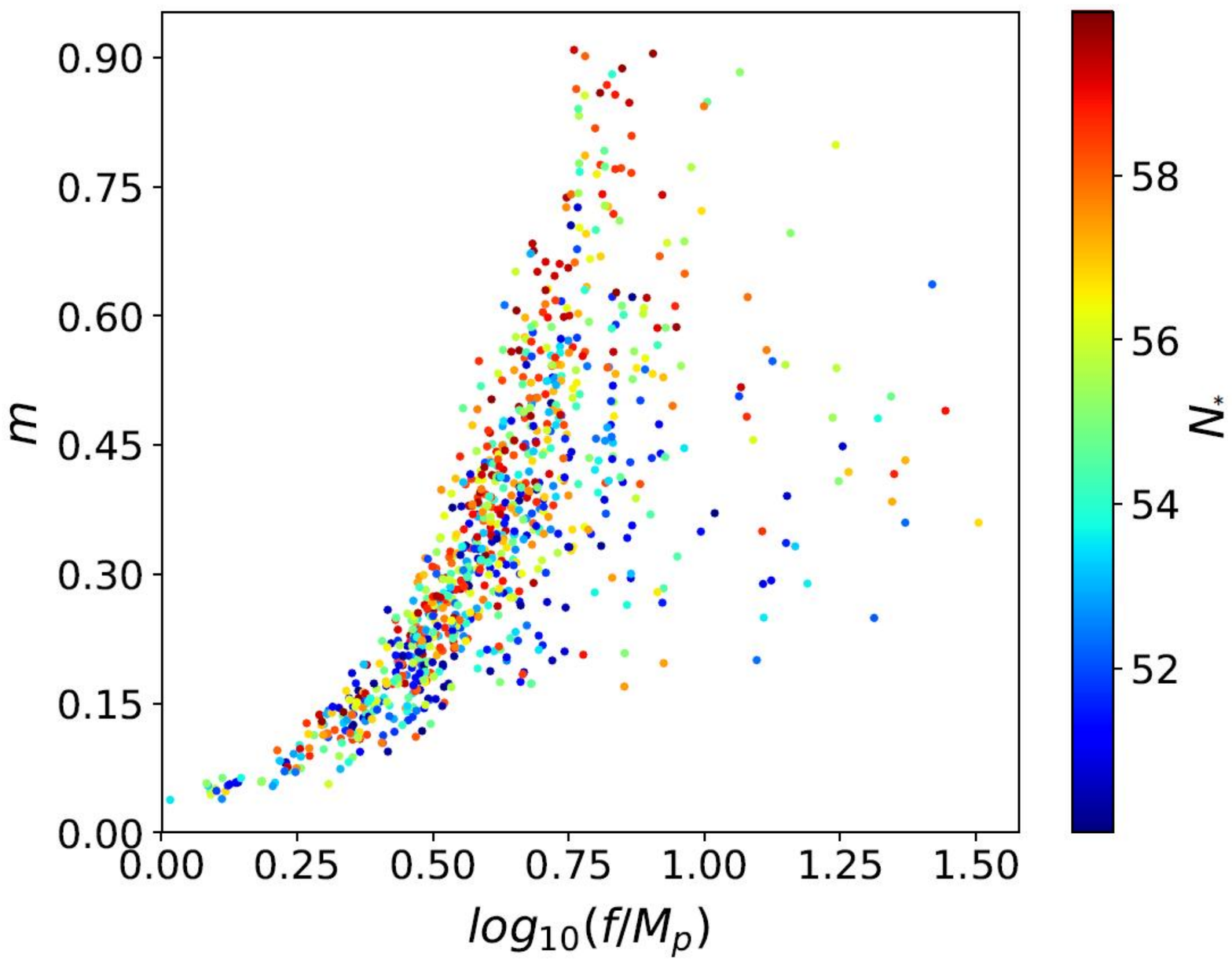}
\caption{\label{fig:parameter}The contour plots for $N_{*}$, $f/M_{p}$ and $m$ in the GNI model from the data of P18+BK15, the values of $N_{*}$ are represented by the color of the points.}
\end{figure}

\begin{figure}[htbp]
\centering
\includegraphics[width=5cm]{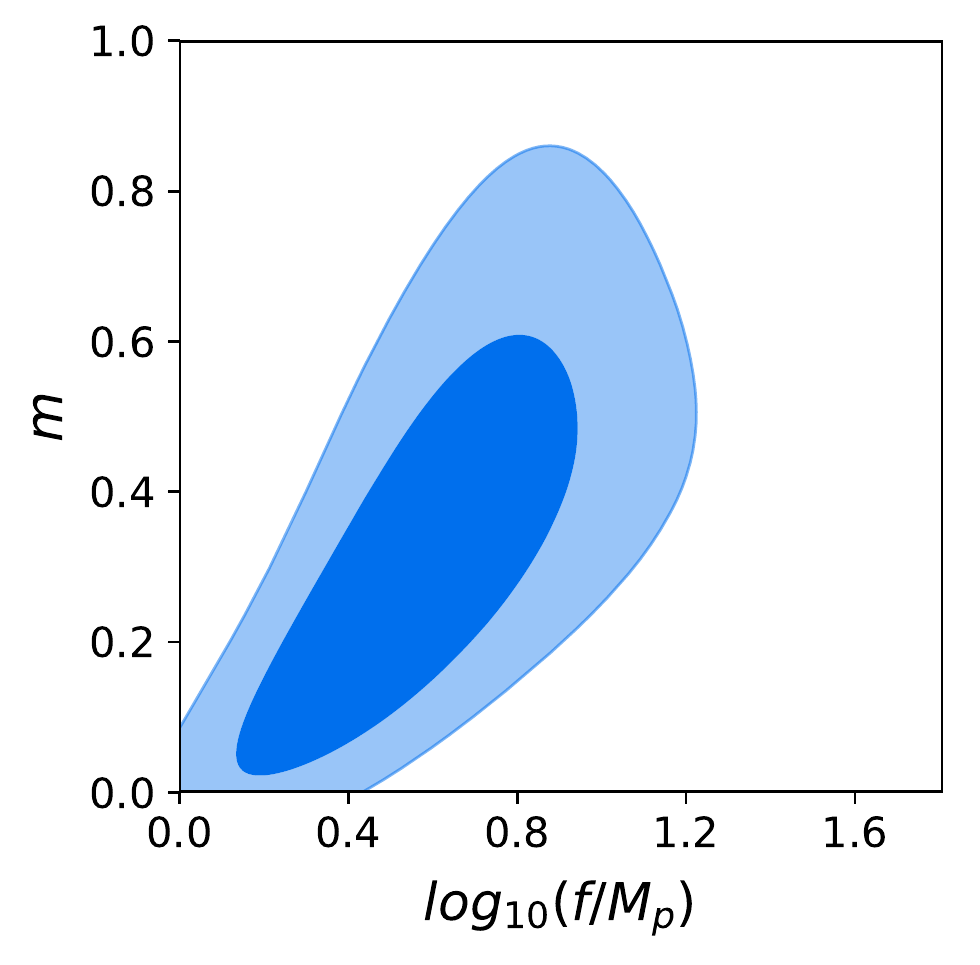}
\caption{\label{fig:parameterfm}The marginalized contour plots for $f/M_{p}$ and $m$ in the GNI model at 68\%CL and 95\%CL from the data of P18+BK15, respectively.}
\end{figure}

Below, $N_{*}$, $f/M_{p}$ and $m$ are all taken as free parameters of the GNI model, the contour plots by means of the public code Cosmomc are shown in Figure~\ref{fig:parameter}.
It shows that the majority of the points locate in the range of $f/M_{p}<10$, and the points corresponding to the small values of $N_{*}$ vanish when the value of $m$ is slightly big.
This result is consistent with our simple predictions in Figure~\ref{fig:nsrf}.
In order to provide a direct sketch, we give the marginalized contour plots for $f/M_{p}$ and $m$ in the usual range $N_{*}\in[50, 60]$ in Figure~\ref{fig:parameterfm}.
The constraints on $f/M_{p}$ and $m$ at $68\%$CL can be read as
\begin{equation}\label{eq:dataf}
\log_{10}(f/M_{p})=0.62^{+0.17}_{-0.18}\mbox{,}
\end{equation}
\begin{equation}\label{eq:datam}
m=0.35^{+0.13}_{-0.23}\mbox{.}
\end{equation}

\begin{figure}[htbp]
\centering
\includegraphics[width=6cm]{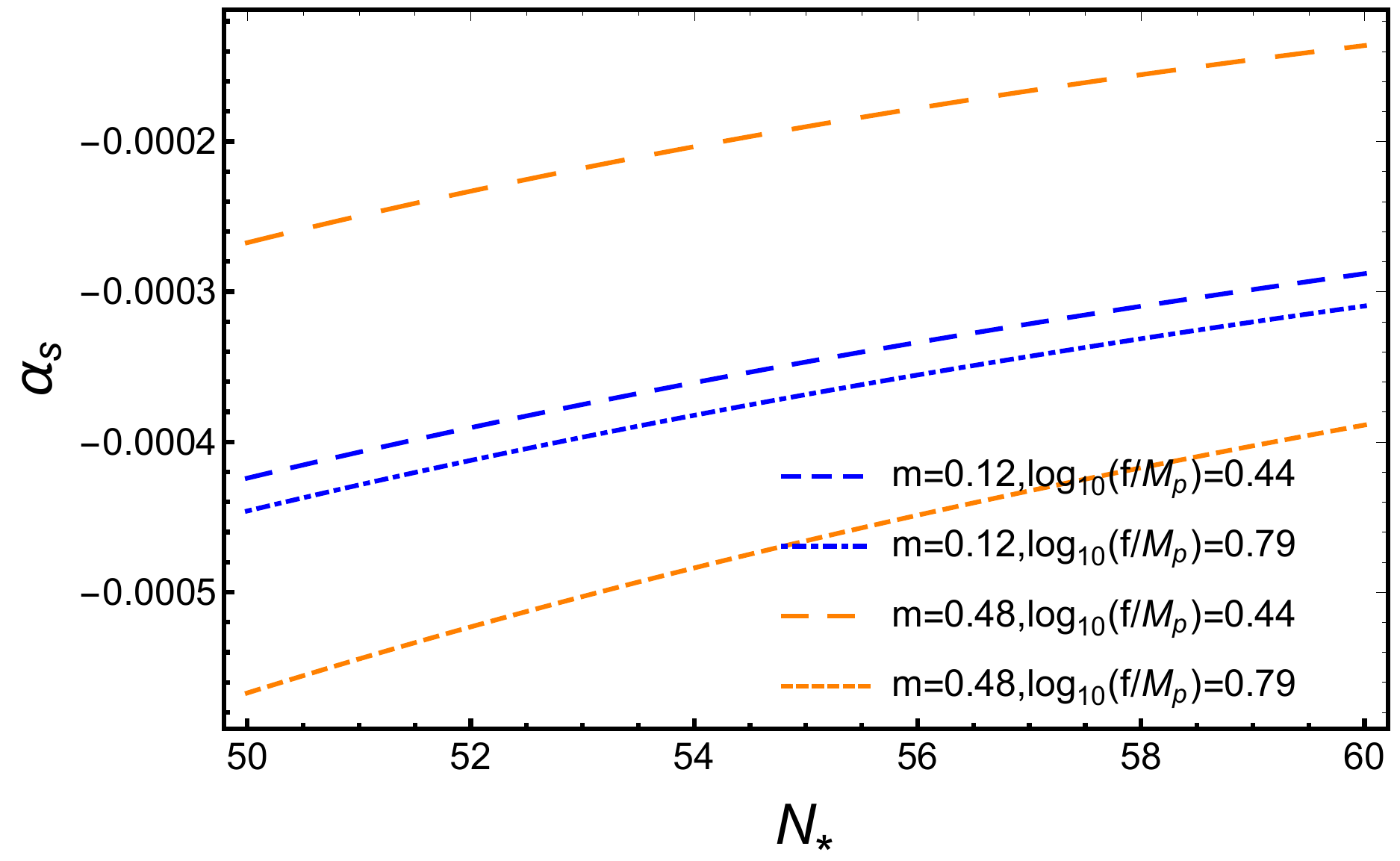}
\caption{\label{fig:run}The running of the scalar spectral index $\alpha_{s}$ with respect to $N_{*}$, the values of $f/M_{p}$ and $m$ are taken as the boundary values of the GNI model given by the data from P18+BK15 at $68\%$ CL.}
\end{figure}

Moreover, we can also calculate the higher-order slow-roll parameter $\xi_{v}^{2}=M_{p}^{4}\frac{V'(\phi)V'''(\phi)}{V(\phi)^{2}}$ and the running of the scalar spectral index $\alpha_{s}\equiv
\frac{dn_{s}}{d\ln k}\simeq 16\epsilon_{v}\eta_{v}-24\epsilon_{v}^{2}-2\xi_{v}^{2}$ of the GNI model as follows:
\begin{equation}\label{eq:ev33}
\xi_{v}^{2}=m^{2}M_{p}^{4}\frac{4f^{2}m^{2}-(3m-1)(2f^{2}+m^{2}M_{p}^{2})e^{\frac{mM_{p}^{2}}{f^{2}}N_{*}}}{(-2f^{3}+(2f^{3}+fm^{2}M_{p}^{2})e^{\frac{mM_{p}^{2}}{f^{2}}N_{*}})^{2}}\mbox{,}
\end{equation}
\begin{equation}\label{eq:dns33}
\alpha_{s}\simeq
\frac{-2m^{2}(m+1)M_{p}^{4}(2f^{2}+m^{2}M_{p}^{2})e^{\frac{mM_{p}^{2}}{f^{2}}N_{*}}}{(-2f^{3}+(2f^{2}+m^{2}M_{p}^{2})fe^{\frac{mM_{p}^{2}}{f^{2}}N_{*}})^{2}}\mbox{.}
\end{equation}
In Figure~\ref{fig:run}, we plot $\alpha_{s}$ with respect to $N_{*}$, the values of $f/M_{p}$ and $m$ are taken as the boundary values of Eqs.~(\ref{eq:dataf})(\ref{eq:datam}).
It shows that the value of $\alpha_{s}$ is negative and its value increases with increasing $N_{*}$ for the fixed values of $f$ and $m$.
And the allowable range of values of $\alpha_{s}$ becomes wide with increasing $m$ when $\log_{10}(f/M_{p})=0.62^{+0.17}_{-0.18}$, $-0.0006\lesssim\alpha_{s}\lesssim-0.0001$ can be found from Figure~\ref{fig:run}.

\begin{figure}[htbp]
\centering
\includegraphics[width=6cm]{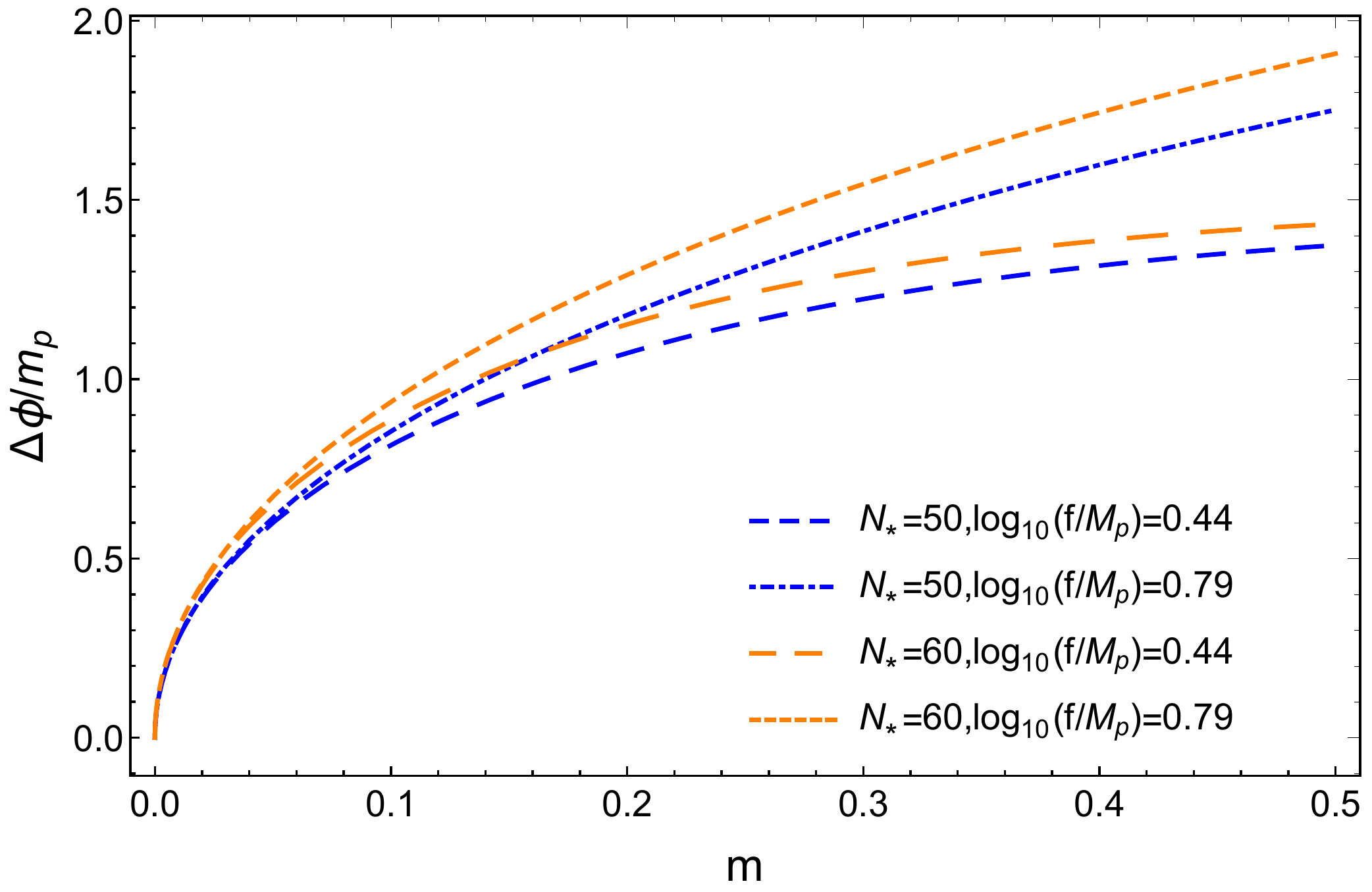}
\caption{\label{fig:res}The excursion $\Delta\phi/m_{p}$ with respect to $m$, the values of $N_{*}$ are taken as $50$, $60$, and the values of $f$ are taken as the boundary values of the GNI model given by the data from P18+BK15 at $68\%$ CL, respectively.}
\end{figure}

In addition, another important issue of single field inflationary model is the so-called trans-Planckian field excurison of the inflaton. This problem appears in the model with large tensor-to-scalar ratio $r$, as indicated by the Lyth bound \cite{Lyth:1996im}, and will spoil the basis of the effective field theory.
In the GNI model considered here, $r$ is constrained to be small, so this will not be a serious problem.
We plot the excursion $\Delta\phi=\mid\phi_{*}-\phi_{end}\mid$ of the inflaton with respect to $m$ in Figure~\ref{fig:res}, and find that the excursion $\Delta\phi$ decreases with lower $m$ and $f$. One can see
from Figure~\ref{fig:res} that with small $f$ and $m$, which fit the observational data, $\Delta\phi$ might be below the Planck mass $m_{p}=1/\sqrt{G}=1.2\times10^{19}$GeV.

\section{\label{sec:level3}The reheating phase of the GNI model}

Following the approaches proposed in Refs.~{\cite{UandY2016,MandR2010,Det2014}}, as for the reheating phase, the starting point is the relation $k=aH$. It leads to:
\begin{eqnarray}\label{eq:aH}
0 &=&\ln\frac{a_{end}}{a_{*}}+\ln\frac{a_{re}}{a_{end}}+\ln\frac{a_{0}}{a_{re}}+\ln\frac{k_{*}}{a_{0}H_{*}} \nonumber\\
  &=&N_{*}+N_{re}+\ln\frac{a_{0}}{a_{re}}+\ln\frac{k_{*}}{a_{0}H_{*}}\mbox{,}
\end{eqnarray}
where the subscript ``$re$'' corresponds to the end of reheating, $a_{0}$ means the present value of scale factor which is equal to $1$, the pivot scale $k_{*}$ is chosen to be 0.05 Mpc$^{-1}$.
Based on the conservation of entropy density $g_{re}a_{re}^{3}T_{re}^{3}=g_{\gamma}a_{0}^{3}T_{\gamma}^{3}+\frac{7}{8}g_{\nu}a_{0}^{3}T_{\nu}^{3}$ and the relationship of temperature $T_{\nu}/T_{\gamma}=(4/11)^{1/3}$, the expression of $a_{re}/a_{0}$ can be written as $a_{re}/a_{0}=(43/11g_{re})^{1/3}/(T_{\gamma}/T_{re})$, $T_{\gamma}=2.7255$ K is a known quantity.
The parameter $g$ with subscripts is the effect number of degrees of freedom, $g_{\gamma}=2$ and $g_{\nu}=6$, $g_{re}$ is assuming as $10^{3}$ for single scalar field inflationary models in keeping with Planck results \cite{Pgroup2016A20,Pgroup2018X,book}.

Considering the energy density of the universe at the end of the reheating $\rho_{re}=g_{re}\frac{\pi^{2}}{30}T_{re}^{4}$ and the continuity equation $\rho_{re}=\rho_{end}\exp[-3(1+w_{re})N_{re}]$, the expression of temperature at the end of reheating can be obtained,
\begin{equation}\label{eq:Tre}
T_{re}=\exp[-\frac{3}{4}(1+w_{re})N_{re}]~(\frac{45V_{end}}{g_{re}\pi^{2}})^{1/4}\mbox{,}
\end{equation}
where $w_{re}$ is regarded as the average EoS during reheating, $V_{end}$ is used to substitute for $\rho_{end}$ \cite{book}. The relation between $V_{end}$ and $\rho_{end}$ can be deduced by taking $\epsilon_{H}=\frac{3}{2}(1+w)=\epsilon_{end}$, and $\epsilon_{end}=1$ is the sign of the end of inflation. $\epsilon_{H}\equiv-\dot{H}/H^{2}$ is the first Hubble hierarchy parameter, in slow-roll approximation, $\epsilon_{H}\simeq\epsilon_{v}$.
For a scalar field, $w\equiv P/\rho=\frac{\frac{1}{2}\dot{\phi}^{2}-V}{\frac{1}{2}\dot{\phi}^{2}+V}$. After a simple calculation, we can get $\rho_{end}\simeq\frac{3}{2}V_{end}$.
Hence, the third term on the righthand in Eq. (\ref{eq:aH}) can be rewritten as
$
\ln\frac{a_{0}}{a_{re}}=\frac{1}{3}\ln\frac{11g_{re}}{43}-\frac{3}{4}(1+w_{re})N_{re}+\frac{1}{4}\ln\frac{45V_{end}}{g_{re}\pi^{2}}-\ln T_{\gamma}
$.
Therefore, $H_{*}$ is the only uncertain quantity in Eq. (\ref{eq:aH}) and it can be fixed by combining Eqs. (\ref{eq:Fsr})(\ref{eq:r}) with the scalar amplitude $A_{s}$ as $A_{s}\simeq H_{*}^{2}/(\pi^{2}M_{p}^{2}r/2)$.

Finally, the expression of the e-folding number $N_{re}$ during reheating can be written as follows:
\begin{eqnarray}\label{eq:Nre}
N_{re}&=&\frac{4}{1-3w_{re}}[-N_{*}-\frac{1}{3}\ln\frac{11g_{re}}{43}-\frac{1}{4}\ln\frac{45V_{end}}{g_{re}\pi^{2}} \nonumber\\
       &&~ -\ln\frac{k_{*}}{T_{\gamma}}+\frac{1}{2}\ln (\pi^{2} M_{p}^{2}(r/2)A_{s})]\mbox{.}
\end{eqnarray}

\subsection{\label{sec:level31}The general reheating phase}

\begin{figure*}[htbp]
\centering
\includegraphics[width=5cm]{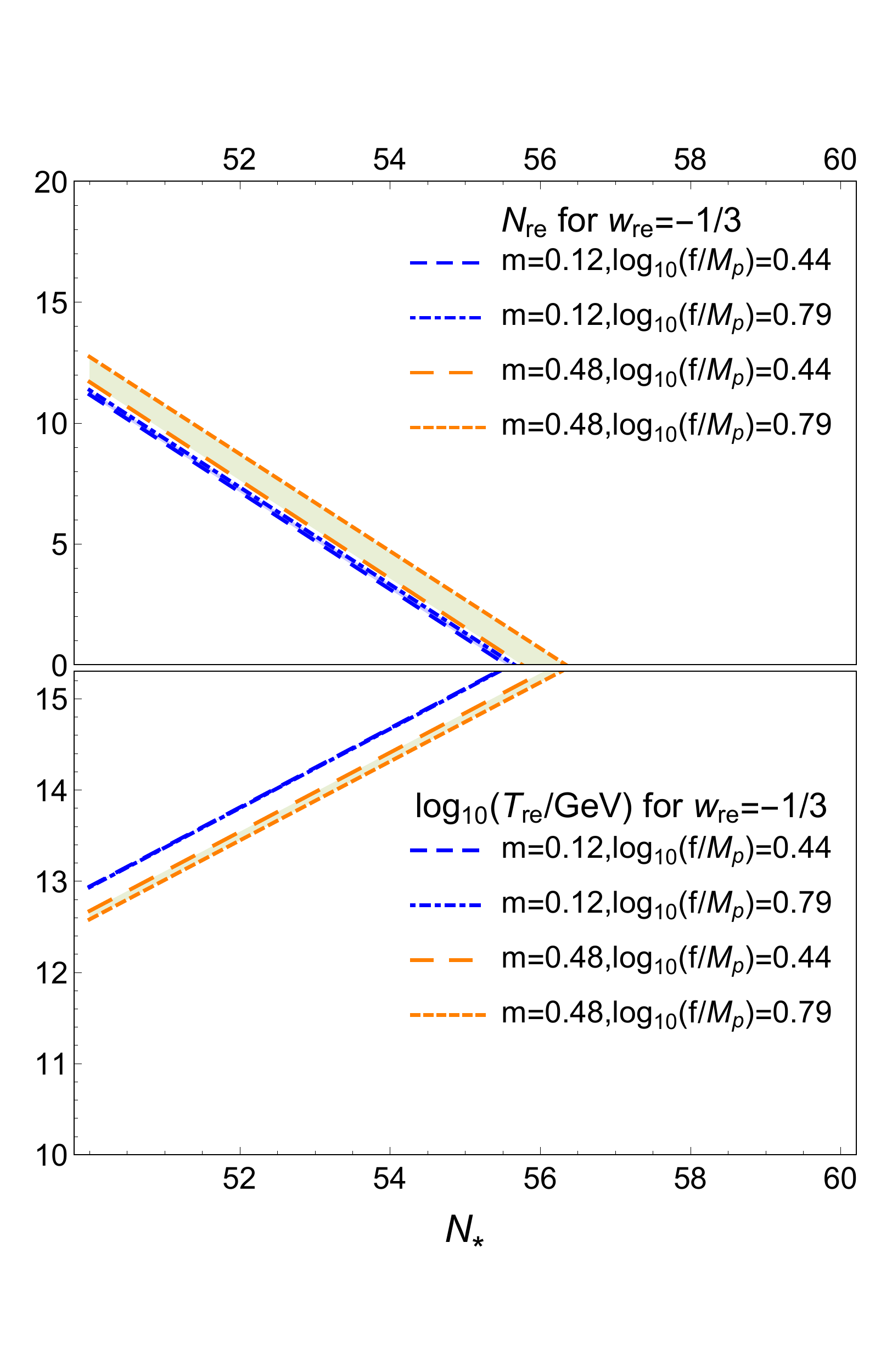}
\includegraphics[width=5cm]{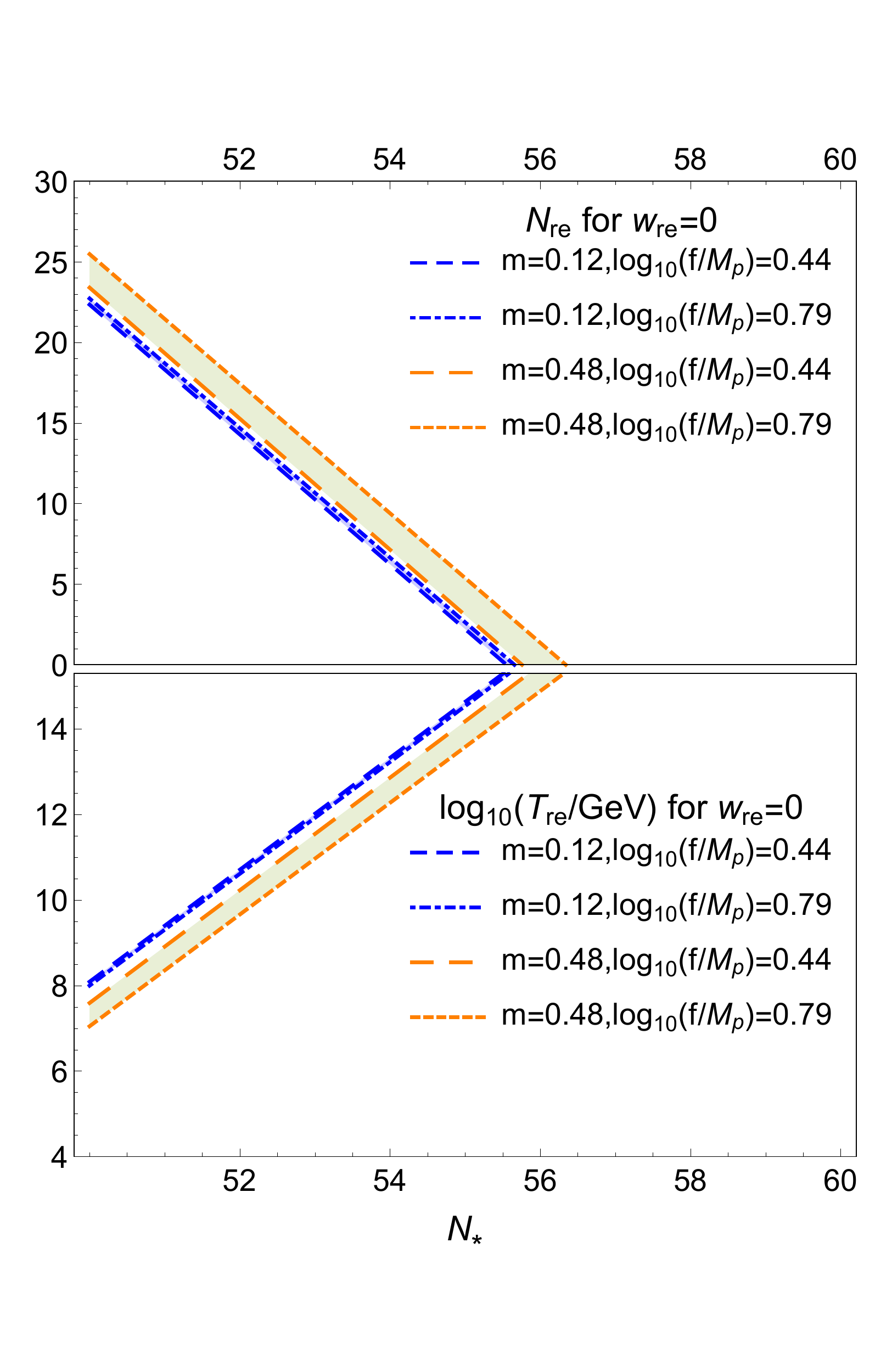}
\includegraphics[width=5cm]{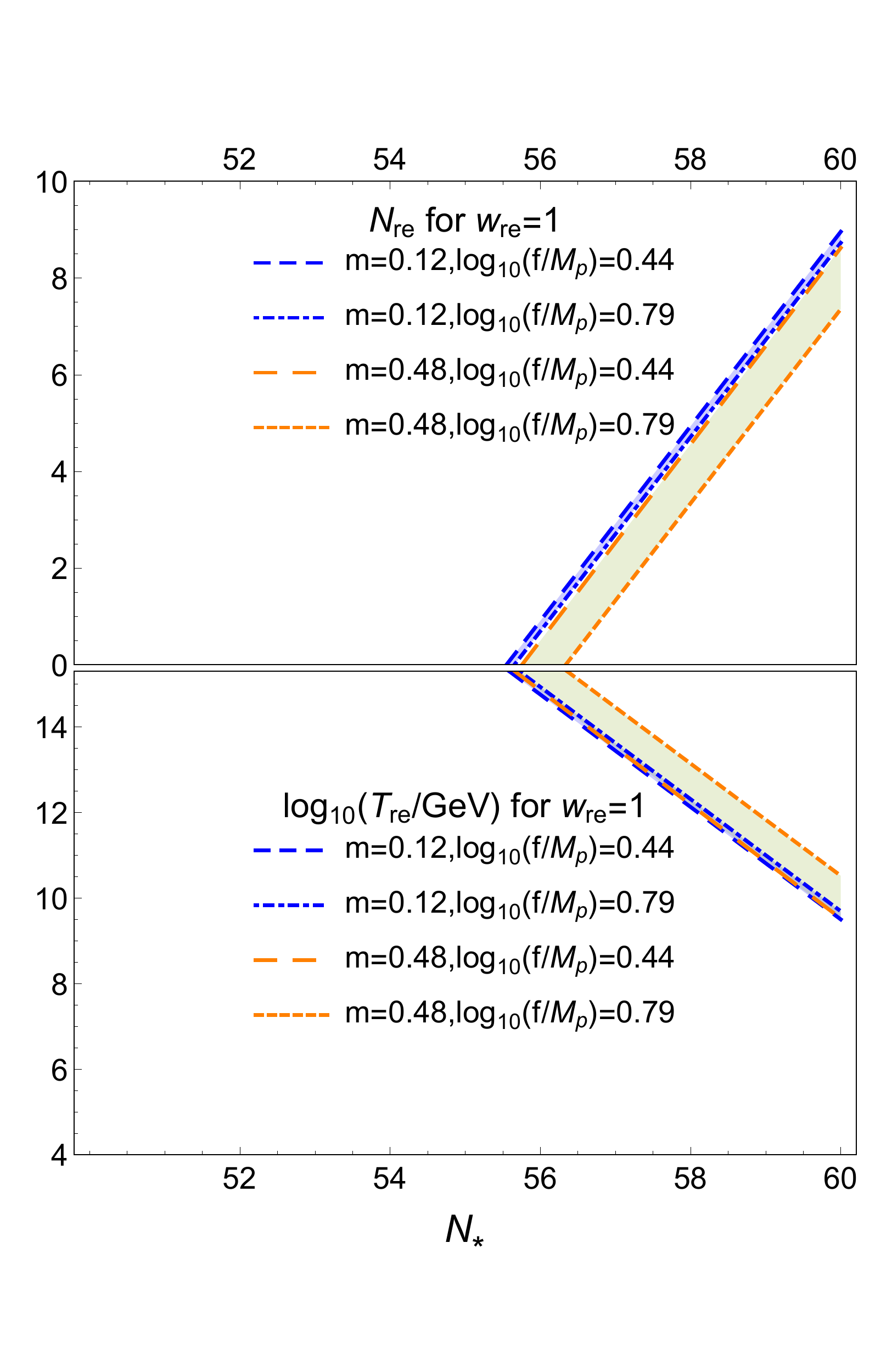}
\caption{The evolving trajectories of the reheating e-folding number $N_{re}$ and the reheating temperature $T_{re}/$ GeV with respect to $N_{*}$ are respectively plotted in the upper and lower panels, $w_{re}$ are taken as $w_{re}=-1/3, 0$ and $1$ in each panel from left to right, the values of $f/M_{p}$ and $m$ are taken as the boundary values of the GNI model given by the data from P18+BK15 at $68\%$ CL.}
\label{fig:reh} 
\end{figure*}

For general reheating epoch, substituting the expressions of $V_{end}$ and $r$ into Eqs.~(\ref{eq:Tre}) and (\ref{eq:Nre}), $N_{re}$ and $T_{re}$ can be obtained for the GNI model.
As we know, the EoS for the scalar field is in the range of $[-1, 1]$, and it should be smaller than $-1/3$ to meet the requirement of accelerating expansion.
Thus, when discussing the reheating phase, the value of $w_{re}$ is usually taken in the range of $-1/3\leq w_{re}\leq1$.
In addition, it is easy to find the denominator of Eq.~(\ref{eq:Nre}) will vanish if $w_{re}=1/3$, which means $w_{re}=1/3$ is the boundary of different evolution of tendencies for reheating parameters.
The evolving trajectories of the reheating e-folding number $N_{re}$ and the reheating temperature $T_{re}$ are illustrated in Figure~\ref{fig:reh}, three different values of $w_{re}$ are considered in each panel as $w_{re}=-1/3, 0$ and $1$.
The values of $f/M_{p}$ and $m$ are taken as the boundary values of Eqs.~(\ref{eq:dataf}) and (\ref{eq:datam}).

It can be seen that the high limit of the reheating temperature $T_{re}$ and the limits of $N_{*}$ can be obtained by the physical condition that the reheating e-folding number $N_{re}$ should be non-negative.
There are the maximum and minimum values of $N_{*}$ for the cases of $w_{re}<1/3$ and $w_{re}>1/3$, respectively, i.e., $N_{*}^{max}=56.34$ and $N_{*}^{min}=55.5$.
Moreover, the ranges of corresponding values of reheating parameters increase first and then decrease with the increasing $w_{re}$.
And we can find that $N_{re}$ and $T_{re}$ are slightly more sensitive to the value of $m$ than the value of $f$.
When $w_{re}<1/3$, $N_{re}$ increases and $T_{re}$ decreases at the fixed value of $m$ (or $f/M_{p}$) with increasing $f/M_{p}$ (or $m$).
When $w_{re}>1/3$, the reheating parameters evolve in the opposite trends.
The concrete results of the corresponding ranges of values of $N_{re}$ and $\log_{10}(T_{re}/$GeV) for $w_{re}=-1/3, 0$ and $1$ in the ranges of $\log_{10}(f/M_{p})=0.62^{+0.17}_{-0.18}$ and $m=0.35^{+0.13}_{-0.23}$ are listed in Table \ref{tab:tre}.
\begin{table}[htbp]
\caption{\label{tab:tre}The corresponding ranges of values of $N_{re}$ and $\log_{10}(T_{re}/$GeV) for $w_{re}=-1/3, 0$ and $1$ in the ranges of $\log_{10}(f/M_{p})=0.62^{+0.17}_{-0.18}$ and $m=0.35^{+0.13}_{-0.23}$.}
\centering
\begin{tabular}{ccc}
\hline \hline
$w_{re}$      &$N_{re}$         &$\log_{10}(T_{re}/$GeV)           \\ \hline
$-1/3$       &$0-12.7$        &$12.6-15.3$           \\
$0$          &$0-25.5$       &$7.0-15.3$                   \\
$1$         &$0-9.0$        &$9.5-15.3$                   \\              \hline \hline
\end{tabular}
\end{table}

\subsection{\label{sec:level32}The two-phase reheating}

Below we consider the reheating scenario as a simple case of two-phase process.
After the end of inflation, scalar field inflaton starts to oscillate and decay into radiation field $\chi$ which is the so-called oscillation phase.
At the equal scale, when the energy density of the oscillation field equals to the one of relativistic particle field, i.e., the expansion Hubble constant $H$ equals to the decay rate $\Gamma$, the universe is going to be dominated by radiation.
Therefore, $H=\Gamma$ is regarded as the sign of completing the simplest two-phase reheating.
Note that, the system is not at thermal equilibrium during the process, and it has gone through a process called thermalization phase.
When the $\phi$ field oscillates around its minimum value, the potential Eq.~(\ref{eq:GNIV}) has an approximate form of $V(\phi)\propto\phi^{2m}$.
For the $\phi^{2m}$ form-like potential, the EoS can be expressed as \cite{UandY2016,Mbook2005}:
\begin{equation}\label{eq:wsc}
w_{sc}=\frac{m-1}{m+1}\mbox{.}
\end{equation}
The reheating e-folding number $N_{re}$ is reconsidered as the sum of two phases, $N_{sc}$ and $N_{th}$.
We know that $N_{sc}=\ln\frac{a_{eq}}{a_{end}}$ and $N_{th}=\ln\frac{a_{re}}{a_{eq}}$, where $a_{eq}$ is the dividing point of the two phases. They can be written as
\begin{equation}\label{eq:Nsc2}
N_{sc}=-\frac{1}{3(1+w_{sc})}\ln\frac{\rho_{eq}}{\rho_{end}}\mbox{,}
\end{equation}
and
\begin{equation}\label{eq:Nth2}
N_{th}=-\frac{1}{4}\ln\frac{\rho_{re}}{\rho_{eq}}\mbox{,}
\end{equation}
where $w_{th}=w_{r}=1/3$ has been adopted in Eq.~(\ref{eq:Nth2}).

Based on the continuity equation, $\rho_{eq}$ can be expressed as
\begin{eqnarray}\label{eq:DenSc}
\rho_{eq}
&=&\frac{3}{2}V_{end}\exp[-3(1+w_{sc})N_{sc}]\mbox{.}
\end{eqnarray}
Finally, Eqs.~(\ref{eq:Nsc2}) and (\ref{eq:Nth2}) can be rewritten as follows:
\begin{eqnarray}\label{eq:NscEnd}
N_{sc}=&&\frac{4}{1-3w_{sc}}[-N_{*}-\frac{1}{3}\ln\frac{11g_{re}}{43}-\frac{1}{4}\ln\frac{45V_{end}}{g_{re}\pi^{2}} \nonumber\\
      &&-\ln\frac{k_{*}}{T_{\gamma}}+\frac{1}{2}\ln\frac{\pi^{2} M_{pl}^{2}rA_{s}}{2}]\mbox{,}
\end{eqnarray}
\begin{equation}\label{eq:Tsc}
T_{re}e^{N_{th}}=\exp[-\frac{3}{4}(1+w_{sc})N_{sc}]~(\frac{45V_{end}}{g_{re}\pi^{2}})^{1/4}\mbox{.}
\end{equation}

\begin{figure}[htbp]
\centering
\includegraphics[width=5cm]{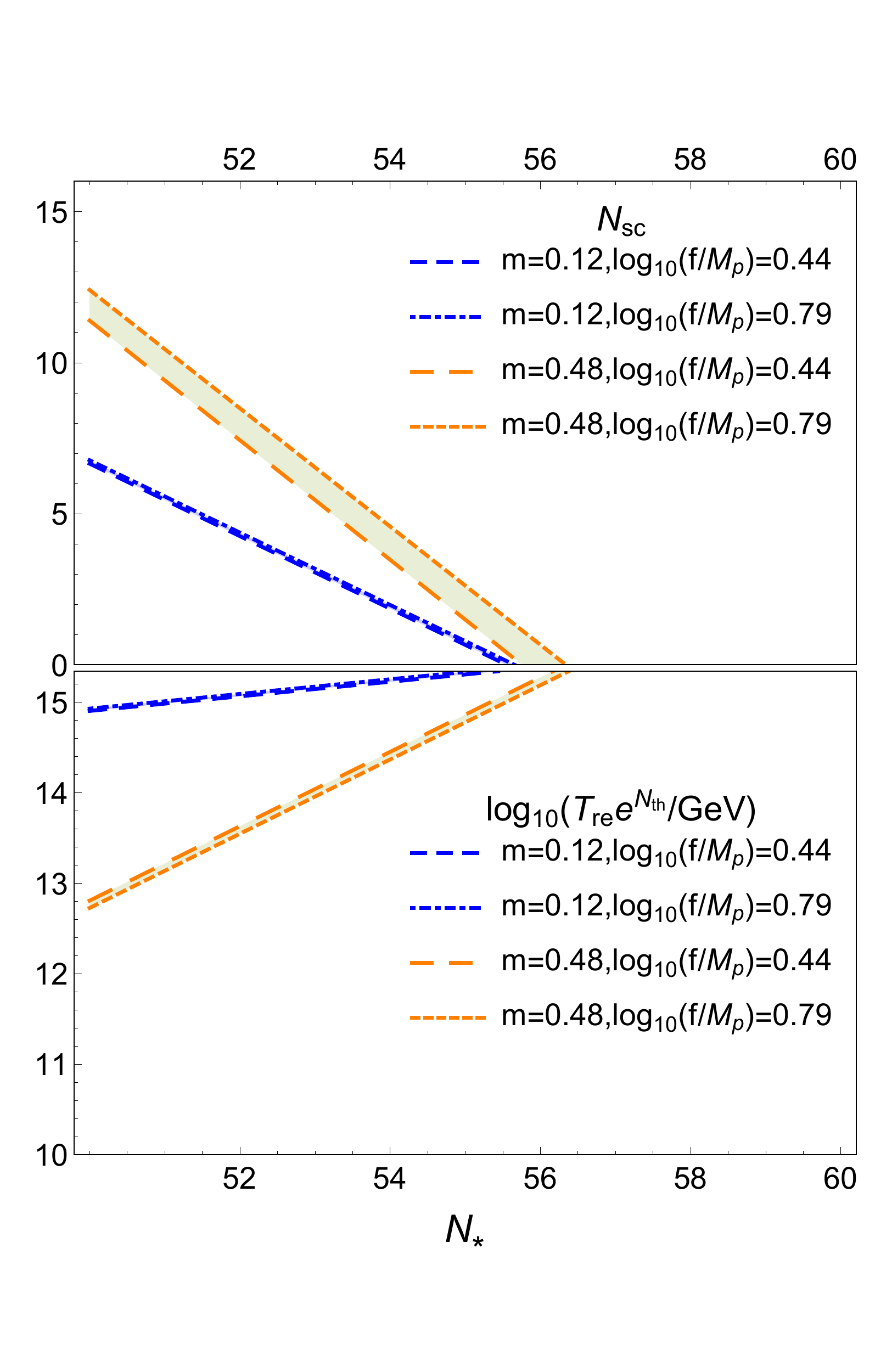}
\caption{The evolutions of $N_{sc}$ and $T_{re}e^{N_{th}}/$GeV with respect to $N_{*}$ in the two-phase reheating scenario are respectively plotted in the upper and lower panels, the values of $f/M_{p}$ and $m$ are taken as the boundary values of the GNI model given by the data from P18+BK15 at $68\%$ CL.}
\label{fig:sc} 
\end{figure}

Figure~\ref{fig:sc} shows the evolutions of oscillation e-folding number $N_{sc}$ and the temperature $T_{re}e^{N_{th}}/$GeV with $N_{*}$.
We find $N_{sc}$ decreases, while $T_{re}e^{N_{th}}$ increases with increasing $N_{*}$ for the fixed values of $f/M_{p}$ and $m$.
Similar to the cases of general reheating phase, $N_{sc}\geq0$ is required in order to make the two-phase reheating meaningful.
The values of $N_{sc}$ and $T_{re}e^{N_{th}}/$GeV are in the ranges of $[0, 12.4]$ and $[12.7, 15.3]$, respectively.

In addition, we consider the elementary decay $\phi\rightarrow\chi\chi$ with the interaction $-g\phi\chi^{2}$, where $g$ respectives the coupling constant.
And following Ref.~\cite{UandY2016}, we take the corresponding decay rate $\Gamma$ as $\Gamma_{\phi\rightarrow\chi\chi}=\frac{g^{2}}{8\pi m_{\phi}}$, where $m_{\phi}$ is the mass of inflaton.
Considering the Friedmann Equation $H^{2}=\frac{\rho_{eq}}{3M_{p}^{2}}$ and the equality of $H=\Gamma$, it can be directly obtained that $(\frac{\rho_{eq}}{3M_{p}^{2}})^{1/2}=\frac{g^{2}}{8\pi m_{\phi}}$.
Substituting Eqs.~(\ref{eq:wsc}) and (\ref{eq:DenSc}) into the above equation, the coupling constant $g$ of the GNI model can be deduced as follows:
\begin{equation}\label{eq:g1}
g=\sqrt{\frac{8\pi m_{\phi}}{M_{p}}}(\frac{V_{end}}{2})^{1/4}\exp [-\frac{3m}{2(m+1)}N_{sc}]\mbox{.}
\end{equation}
Utilizing $V_{end}\sim 2^{1-m}\Lambda^{4}(\frac{mM_{p}}{f})^{2m}$, the effective mass of inflaton $m_{\phi}^{2}\sim m\frac{\Lambda^{4}}{f^{2}}$ at vacuum and Eq.~(\ref{eq:NscEnd}), $g$ can be expressed in terms of ($N_{*}$, $f$, $m$).
Therefore, we can give the evolution of the coupling constant $g$ to realize a successful simplest two-phase reheating scenario.

\begin{figure}[htbp]
\centering
\includegraphics[width=6cm]{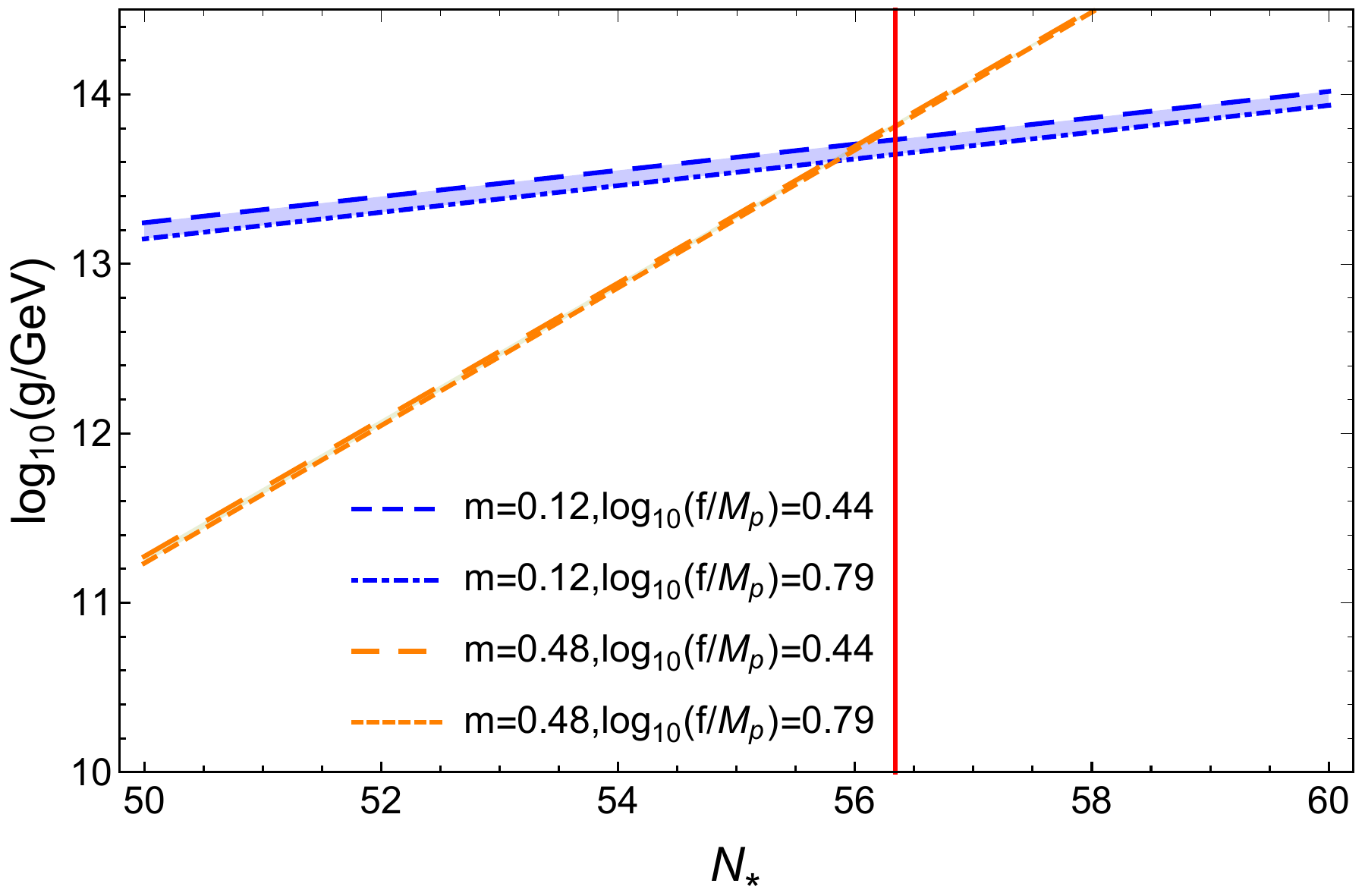}
\caption{The evolution trajectories of the coupling constant $g/$GeV with respect to $N_{*}$, the values of $f/M_{p}$ and $m$ are taken as the boundary values of the GNI model given by the data from P18+BK15 at $68\%$ CL.}
\label{fig:gcontour} 
\end{figure}
In Figure~\ref{fig:gcontour}, we plot the coupling constant $g$ with respect to $N_{*}$, the solid (red) line corresponds to $N_{*}=56.34$ from the requirement of $N_{sc}\geq0$.
From Figure~\ref{fig:gcontour}, we find that the value of $g$ increases with $N_{*}$ for the fixed values of $f/M_{p}$ and $m$.
Besides, it shows that the value of $g$ is more sensitive to the value of $m$ than the value of $f$.
And $g$ decreases with increasing $f$ for the fixed value of $N_{*}$ and $m$.
However, it decreases first and then increases with increasing $m$ for the fixed value of $N_{*}$ and $f$, there are turning points for $\log_{10}(f/M_{p})=0.79$ and $0.44$ as $N_{*}=55.8$ and $56.0$, respectively.
The minimum value of $g$ to complete the two-phase reheating is at the order of magnitude of $10^{11}$GeV.

\section{\label{sec:level14}conclusions}

In summary, in the paper we have in detail investigated the generalized natural inflationary (GNI) model according to the latest observational data from Planck 2018 plus BK15.
As we discussed above, the constraints on the observables $n_{s}$ and $r$ for $\Lambda$CDM $+r$ model by means of the available Cosmomc code are given as $n_{s}=0.9659\pm0.0044$ at $68\%$CL and $r<0.0623$ at $95\%$CL (Planck TT,TE,EE+lowE+lensing+BK15).

For the GNI model, $n_{s}$ and $r$ have been expressed by $N_{*}$, $f/M_{p}$ and $m$, thus the allowable parameter space of $(f/M_{p}, m)$ in the range of $N_{*}\in[50, 60]$ have been obtained as $\log_{10}(f/M_{p})=0.62^{+0.17}_{-0.18}$ and $m=0.35^{+0.13}_{-0.23}$ at $68\%$CL.
The general reheating parameters in the GNI model are the functions of $N_{*}$, $f$, $m$ and $w_{re}$.
Thus, we have obtained the evolutions of $N_{re}$ and $\log_{10}(T_{re}/$GeV) with $N_{*}$ for $w_{re}=-1/3, 0$ and $1$, respectively.
The corresponding ranges of values have been shown as $N_{re}\leq25.5$ and $7\leq\log_{10}(T_{re}/$GeV)$\lesssim15$ within the obtained allowable parameter space when $w_{re}=0$.
As for the two-phase reheating, the parameters are the functions of $N_{*}$, $f$ and $m$, the values of $N_{sc}$ and $T_{re}e^{N_{th}}/$GeV are in the ranges of $[0, 12.4]$ and $[12.7, 15.3]$, and the values of the coupling constant $g/GeV$ is in the range of $11.2\leq\log_{10}(g/GeV)\leq13.8$  within the allowable parameter space.
It follows that the parameter $m$ has the significant effect on the behaviors of the GNI model.

\section*{Acknowledgments}

We thank Dr. Xue Zhang for her helpful discussion on the Monte Carlo method.
This work is supported by the National Natural Science Foundation of China (Grant Nos. 11575075, 11705079 and 11865012).

\end{document}